\documentstyle [fleqn,12pt]{article}
\begin{document}
\baselineskip .75cm 
\begin{titlepage}
\title{ \bf Self-consistent quasi-particle model for relativistic plasma}      
\author{Vishnu M. Bannur \\
{\it Department of Physics}, \\  
{\it University of Calicut, Kerala-673 635, India.} }
\maketitle
\begin{abstract} 

Relativistic plasma with radiation at thermodynamic equilibrium is a 
general system of interest in astrophysics and high energy physics.  
We develop a new self-consistent quasi-particle model for such a system  
to  take account of collective behaviour of plasma and  
thermodynamic properties are derived.
It is applied to electrodynamic plasma and quark gluon plasma 
and compared with existing results.   
\end{abstract}
\vspace{1cm}

\end{titlepage}
\section{Introduction :}

A typical electrodynamic plasma of interest in astrophysics \cite{me.1}, 
consists of electron, positron and photons 
($e$, $e^+$, $\gamma$) at thermodynamic 
equilibrium. Since it is a plasma medium, individual particle properties 
are modified by collective effects of plasma. One way to take account 
of this effect, is to use quasi-particle model. Just like in Debye theory 
of specific heat or theory of liquid helium etc. thermal properties 
of the medium may be viewed as a result of thermal excitations or 
quasi-particles, like plasmons and dressed photons as a result of 
quantization of plasma waves and electromagnetic waves in plasma. The 
standard procedure is to obtain the classical dispersion relations for 
plasma or electromagnetic waves in plasma and on their quantization, we 
get quasi-particles, namely quasi-fermions corresponding to  $e$, $e^+$ 
and quasi-bosons like dressed photons. We study the statistical mechanics 
and thermodynamics of such a system of quasi-particles. One such study 
was attempted by Medvedev \cite{me.1} which we modify and correct it 
to get the present model and then generalize it to quark gluon plasma (QGP). 

QGP is a plasma made up of quarks and gluons \cite{qgp.1},
 governed by strong interaction 
called quantum chromodynamics (QCD). It is similar to ($e$, $e^+$, $\gamma$) 
system with electrons (positrons) are replaced by quarks (antiquarks) and 
photons by gluons. QED (quantum electrodynamics) is replaced by QCD. 
QGP also exhibits collective behaviour and hence the thermodynamic 
properties of QGP are modified and may be studied using quasi-particle 
model. Of course, this kind of study was first attempted by Peshier 
{\it et. al.} \cite{pe.1} and latter it was modified with corrections 
and studied various 
groups \cite{pe.2}. However, initially, these models were thermodynamically 
incosistent \cite{go.1} and also not self-consistent in calculating the 
thermodynamics (TD) and quasi-particle properties. Few attempts were done to 
correct these inconsistency \cite{go.1,pe.2} by reformulating statistical 
mechanics (SM). However, in Ref. \cite{ba.1}, we have developed TD consistent 
model of QGP without reformulating SM.  
Here, in this new model, all previous problems of quasi-particle model, 
namely TD inconsistency and self-consistency, are solved.   
  
\section{Quasi-particle model of plasma :}
Both the systems we discussed are highly relativistic and hence we 
develop here quasi-particle model for such ultra-relativistic systems. 
We assume that the collective excitations of plasma 
leads to a system of non-interacting quasi-fermions and 
quasi-bosons, obeying Fermi and Bose statistics respectively. Following the 
standard statistical mechanics \cite{pa.1}, 
density of quasi-particles may be written as, 
\begin{equation}
n = \frac{1}{V}\,\sum_k \frac{1}{z^{-1}\, e^{\beta \epsilon_k} \mp 1}  
\rightarrow \frac{g_f}{2 \pi^2}\,\int_0^{\infty} dk\,k^2\, \frac{1}{
z^{-1}\, e^{\beta \epsilon_k} \mp 1} \,\,, 
\end{equation}
where $z$ is the fugacity, $\mp$ refers to Bosons and Fermions. 
$g_f$ is the degenarcy associated with the internal degrees of freedom. Here   
$\epsilon_k$ is the energy of quasi-particle which may be obtained from 
classical approximate dispersion relation as, 
\begin{equation}
\epsilon_k = \hbar \sqrt{k^2 c^2 + \omega_p^2}\,\,,  
\end{equation}
for Bosons and 
\begin{equation}
\epsilon_k = \hbar \sqrt{k^2 c^2 + \omega_p^2}\,\,,  
\end{equation}
for Fermions. These forms of energy-momentum relations are widely used 
in quasi-particle models of QGP \cite{pe.1,pe.2} with $\omega_p^2$ 
replaced by temperature dependent masses, which they obtain from 
the finite temperature field theory calculations in ideal thermal bath.  
The general expressions for $\epsilon_k$ are very complicated even at 
high momentum limit and following Medvedev \cite{me.1}, we approximate them 
to above simpler equations with an error of about $3\%$. $c$ is the 
speed of light and $\hbar$ is the Planck constant.  
$\omega_p$ is the plasma frequency, given by, 
\begin{equation}
\omega_p^2 = \frac{8\,\pi\,e^2\,n_e\,c^2}{3\,T} 
\equiv a \frac{n_e}{T}
\,\,, \end{equation} 
for ($e$, $e^+$, $\gamma$) system. $e$ is the charge and $n_e$ 
the electron 
density which is also equal to positron density for a system with 
chemical potential equal to zero. In quasi-particle models, by definition, 
electron density is same as that of quasi-electrons. 
$T$ is the temperature of the system.  

\section{($e$, $e^+$, $\gamma$) system:}  
 
Let us first consider ($e$, $e^+$, $\gamma$) system with chemical 
potential zero, or $z=1$, and the density of 
quasi-electrons is, 
\begin{equation}
n_e = \frac{g_e}{2 \pi^2}\,\int_0^{\infty} dk\,k^2\, \frac{1}{
 e^{\beta \hbar \sqrt{k^2 c^2 + a\,\frac{n_e}{T}}} + 1} \,\,, 
\label{eq:ne0} \end{equation}
which may be rewritten as 
\begin{equation}
n_e = \frac{g_e}{2 \pi^2}\,(\frac{T}{\hbar\,c})^3\,
\int_0^{\infty} dx\,x^2\, \frac{1}{
 e^{\sqrt{x^2 + a\,\hbar^2\,\frac{n_e}{T^3}}} + 1} \,\,.  
\label{eq:ne} \end{equation} 
The density of positrons is same as that of electrons. 
Similarly, the density of quasi-photons may be written as,   
\begin{equation}
n_{\gamma} = \frac{g_{\gamma}}{2 \pi^2}\,(\frac{T}{\hbar\,c})^3\,
\int_0^{\infty} dx\,x^2\, \frac{1}{
 e^{\sqrt{x^2 + a\,\hbar^2\,\frac{n_e}{T^3}}} - 1} \,\,, 
\label{eq:np} \end{equation} 
where the fugacity is one for photons. 
These equations need to be solved 
self-consistently because $n_e$ which is to be determined is inside 
the integral through $\omega_p$. Redefining the variables, 
the final equation to be solved self-consistently is,  
\begin{equation}
f_e^2 = \int_0^{\infty} dx\,x^2\, \frac{1}{
 e^{\sqrt{x^2 + \bar{a}^2\,f_e^2 }} + 1} \,
= \bar{a}^2 \,f_e^2 \,\sum_{l=1}^{\infty} \,\frac{(-1)^{(l-1)}}{l}\,
K_2 (\bar{a}\,l\,f_e) \,\,,   
\label{eq:fe0} \end{equation} 
where 
\[ \bar{a}^2 \equiv \frac{4\,g_e}{3\,\pi} \,\alpha \,,\] 
and 
\[ f_e^2 \equiv \frac{2\,\pi^2\,(\hbar\,c)^3}{g_e}\,\frac{n_e}{T^3}\,.\] 
$\alpha$ is the fine structure constant and $K$ is the 
modified Bessel function. 
Once we know $n_e$ or $f_e^2$, we can obtain photon density from the 
relation,  
\begin{equation}
n_{\gamma} = \frac{g_{\gamma}}{2 \pi^2}\,(\frac{T}{\hbar\,c})^3\,
\bar{a}^2 \,f_e^2\,\sum_{l=1}^{\infty} \,\frac{1}{l}\, 
K_2 (\bar{a}\,l\,f_e) \,\,, 
\end{equation}  
which follows from the Eq. (\ref{eq:np}).  Similarly,  
energy densities are given by,  `
\begin{equation}
\varepsilon_e = \frac{g_e}{2 \pi^2}\,\frac{T^4}{(\hbar\,c)^3}\,
\int_0^{\infty} dx\,x^2\, \frac{\sqrt{x^2 + \bar{a}^2 \, f_e^2 }}{
 e^{\sqrt{x^2 + \bar{a}^2 \, f_e^2 }} + 1} \,\,, 
 \end{equation}
or 
\begin{equation}
\varepsilon_e = \frac{g_e}{2\,\pi^2}\,\frac{T^4}{(\hbar\,c)^3} \sum_{l=1}^{\infty} 
\frac{(-1)^{l-1}}{l^4} \, \left. \left[ (\bar{a} f_e l)^3 
K_1 (\bar{a} f_e l) +  3\, (\bar{a} f_e l)^2 K_2 (\bar{a} f_e l) 
\right] \right)\,\,, 
\end{equation}
in terms of modified Bessel functions $K$, for electrons and  
\begin{equation}
\varepsilon_{\gamma} = \frac{g_{\gamma}}{2 \pi^2}\,\frac{T^4}{(\hbar\,c)^3}\,
\int_0^{\infty} dx\,x^2\,  \frac{\sqrt{x^2 + \bar{a}^2 \, f_e^2 }}{
 e^{\sqrt{x^2 + \bar{a}^2 \, f_e^2 }} - 1} \,\,, 
 \end{equation} 
or 
\begin{equation}
\varepsilon_{\gamma} = \frac{g_{\gamma}}{2\,\pi^2}\,\frac{T^4}{(\hbar\,c)^3} 
\sum_{l=1}^{\infty} \frac{1}{l^4} \, 
\left. \left[ (\bar{a} f_e l)^3 
K_1 (\bar{a} f_e l) +  3\, (\bar{a} f_e l)^2 K_2 (\bar{a} f_e l) 
\right] \right)\,\,, 
\end{equation}
for photons. The blackbody Planck's distribution may be easily read from 
above equations as   
\begin{equation}
d\varepsilon_{\gamma} (x) = \frac{g_{\gamma}}{2 \pi^2}\,\frac{T^4}
{(\hbar\,c)^3}\,x^2\,  \frac{\sqrt{x^2 - \bar{a}^2 \, f_e^2 }}{
 e^x - 1} \,dx \,\,, 
 \end{equation} 
where $x \equiv \hbar \,\omega /T$. Thus the Planck's distribution is 
modified for the non-zero value of $\bar{a}$ due to plasma with a cutoff 
frequency related to plasma frequency.   

On taking a limit $a \rightarrow 0$ in  
Eq. (\ref{eq:ne}, \ref{eq:np}), we get 
\[n_e = 2\,\eta (3)\, \frac{g_e}{2\,\pi^2}\,(\frac{T}{\hbar\,c})^3\,,\]
and 
\[n_{\gamma} = 2\,\zeta (3)\, \frac{g_{\gamma}}{2\,\pi^2}\,
(\frac{T}{\hbar\,c})^3\,,\]
where $\zeta (3)$ is a 
Riemann zeta function and has a value approximately $1.2$ and $\eta(3)$ 
is related to $\zeta(3)$ as $\eta(3) = \frac{3}{4}\,\zeta(3)$. 
Hence the $n_e$ and $n_{\gamma}$ are related as 
$n_e = \frac{3}{4}\,n_{\gamma}$ as expected for ideal 
gas. Note that in the earlier calculations, \cite{me.1}, the starting point 
of the formalism is  $n_e = \frac{7}{8} \,n_{\gamma}$, which is not right and 
such a relation is for energy densities. Secondly, 
it is an ideal gas relation which need not be true for non-ideal system 
that we are discussing. Here in our model we don't use such relations between 
$n_e$ and $n_{\gamma}$, rather both of them are calculated self-consistently 
by solving the coupled integral equations, Eq. (\ref{eq:ne}, \ref{eq:np}),    

Other thermodynamic functions, like pressure may be obtained from 
the thermodynamic relation 
\begin{equation}
\varepsilon = T \frac{\partial P}{\partial T} - P\,, \label{eq:ep}    
\end{equation}
which on integration gives $P = \frac{1}{3}\, \varepsilon$, 
same as that of ideal 
relativistic gas. Again this result also differs from Ref. \cite{me.1} where 
their expression for pressure is not valid for massive particles. 
It is interesting to look at the  
expression for plasma frequency,
\begin{equation}
\omega_p^2 = \frac{4}{3\,\pi}\,g_e\,\alpha\,f_e^2\,(\frac{T}{\hbar})^2 
   \approx \frac{\zeta (3)}{\pi^2} \, e^2\,T^2 \,, 
\end{equation}
where we took $f_e^2 \approx 2\,\eta (3) = \frac{3}{2}\,\zeta (3)$.   
It is very close to $ \frac{1}{9}\,e^2\,T^2$, obtained from the 
finite temperature field theory calculations, but note that it is not 
exactly equal because of our self-consistent calculations.  

\section{Quark gluon plasma:} 

From above discussion on ($e$, $e^+$, $\gamma$) system. it is clear, 
in contrary to earlier work \cite{pe.1,pe.2,go.1}, 
one need to study QGP self-consistently 
in quasi-particle model. QGP is expected to form in ultra-relativistic 
heavy ion collisions and may exist in astropysical objects like 
neutron star, early universe etc. \cite{qgp.1}. 
The lattice gauge theory (LGT) simulation 
of QCD predicts such matter, but seems to be non-ideal plasma near 
to the transition temperature from hadrons to QGP. To explain this behaviour 
many phenomenological models, based on QCD and plasma, have been proposed 
and quasi-particle model is one of them. As we discussed in the introduction, 
many of quasi-particle models \cite{pe.1,pe.2,go.1} lack self-consistency 
in calculating the quasi-particle mass and the thermodynamic properties.  
 
Let us consider, as an example, gluon plasma. The expression for gluon 
density, which is to be solved self-consistently, reduces to  
\begin{equation}
1 = \bar{a}^2 \,\sum_{l=1}^{\infty} \,\frac{1}{l}\,
K_2 (\bar{a}\,l\,f_g) \,\,, 
\end{equation}  
where $f_g$ is similar to $f_{\gamma}$ discussed earlier. Once we know 
$f_g$, energy density may be evaluated from the relation,   
\begin{equation}
\varepsilon_g = \frac{g_g\,T^4}{2\,\pi^2}\, 
\sum_{l=1}^{\infty} 
\frac{1}{l^4} \, \left. \left[ (\bar{a} f_g l)^3 
K_1 (\bar{a} f_g l) +  3\, (\bar{a} f_g l)^2 K_2 (\bar{a} f_g l) 
\right] \right)\,\,, 
\end{equation}
expressed in natural units and $g_g = 16$ for gluons. 
Note that, in QGP, gluons also carry charge and hence exhibits plasma 
oscillations even in the absence of quarks. Therefore, in pure gluon 
plasma, plasma frequency depends on gluon density. In QGP, it depends 
both on quark density and gluon density. Hence for gluon plasma we take 
$\omega_p^2 = a\,n_g /T$ with $a \equiv a_0\,\alpha_s$ such that 
$\bar{a}^2 = \frac{a_0\,\alpha_s\,g_g}{2\,\pi^2}$. $\alpha_s$ is the 
QCD running coupling constant. $a_0$ is a constant, equal to $8\,\pi /3$ for  
($e$, $e^+$, $\gamma$) system, but here we determine it by demanding 
that $\omega_p^2 \rightarrow \frac{1}{3}\, g^2\,T^2$ as $T \rightarrow \infty$, 
a perturbative finite temperature field theory result. 
So we get $a_0 \approx 2.15$.   
For constant $\bar{a}$ or $\alpha_s$, 
$\varepsilon_g \propto T^4$ and hence an ideal equation of state (EoS). 
However, LGT results show a modification of $T^4$ law near the transition 
temperature $T_c$ and they actually used temperature dependent 
2-loop order $\alpha_s (T)$. Hence, we also consider similar expression 
to model $\alpha_s (T)$, given by,
\begin{equation} \alpha_s (T) = \frac{6 \pi}{(33-2 n_f) \ln (T/\Lambda_T)}
\left( 1 - \frac{3 (153 - 19 n_f)}{(33 - 2 n_f)^2}
\frac{\ln (2 \ln (T/\Lambda_T))}{\ln (T/\Lambda_T)}
\right)  \label{eq:ls} \;, \end{equation}
where $\Lambda_T$ is a parameter related to QCD scale parameter. $n_f$, 
the number of flavors, is zero in pure gluon plasma to be compared 
with LGT results. $\Lambda_T$ is the only parameter of our model to be 
found by fitting with LGT result. 
Once we know $\varepsilon_g (T)$, it is stright forward to get pressure 
$P(T)$ by integrating the thermodynamic relation Eq. (\ref{eq:ep}) and 
then all other thermodynamic quantities may be obtained.  
      
\section{Results :} 

 For ($e$, $e^+$, $\gamma$) system we recalculated various thermodynamical 
quantities, reported in Ref. \cite{me.1}, using our model with proper 
corrections. The departure of these quantities from that of ideal system 
is too small to be noticed and tabulated in Table 1. Planck's distribution 
in plasma is plotted  in Fig. 1 along with the Planck's distribution in the 
absence of plasma. Qualitatively similar results as in Ref. \cite{me.1} with 
cutoff in the distribution etc. But, note that, the numerical values differ 
and our values listed in Table 1 with plasma are smaller than without 
plasma. Where as in Ref. \cite{me.1} just the opposite, seems to be a 
numerical error. 

When we apply our self-consistent quasi-particle model to QGP, as an exmple 
gluon plasma, we can explain very nicely lattice gauge theory (LGT) 
\cite{ka.1} results as shown in Fig. 2, 
using a single parameter $t_0 \equiv \Lambda_T / T_c = .82$. 
The self-consistently calculated plasma frequency is 
$\frac{1}{3.8}$ $g^2$ $T^2$ at $T = 5\,T_c$, instead of $\frac{1}{2}\,g^2\,T^2$ 
used in all other quasi-particle models \cite{pe.2}.    
It is interesting to notice that it is close to the value we used in 
our earlier thermodynamically consistent 
quasi-particle model $\cite{ba.1}$, $\frac{1}{3}\,g^2\,T^2$.    

\section{Conclusions :}

We formulated a new self-consistent quasi-particle model to describe the 
thermodynamics (TD) of relativistic plasma, like ($e$, $e^+$, $\gamma$) and 
QGP. Basic idea is that because of the collective behaviour of plasma, 
TD of such a system may obtained by studying the TD of quasi-particles 
which are thermally excited quanta of plasma and electromagnetic waves 
in plasma. This is equivalent to a system of bosons and fermions with 
mass propotional to plasma frequency. Plasma frequency depends on the 
density, a TD quantity, which we want to find out and hence a 
self-consistent problem to be solved. Note that, in all earlier 
quasi-particle models of QGP \cite{pe.1, pe.2}, 
this self-consistency is not taken into 
account. In addition they have problems of TD inconsistencies etc. and 
as a result of both problems, they need more than two parameters to fit 
the LGT results. Further extension of this model to QGP with finite number 
of quark flavors, with and without chemical potential, may be future 
interesting problem. 

As long as ($e$, $e^+$, $\gamma$) system, studied earlier by Medvedev 
\cite{me.1}, is concerned, we modified his result with corrections 
in our new approach and obtained qualitatively the same results, 
but the numerical values and the sign of the modifications due to 
plasma are different.


\newpage
\begin{figure}
\caption { The normalized Planck's distribution $u(x)$ with plasma 
(continuous line) and without plasma (dashed line) as a function of  
$x \equiv \hbar \omega / T$. } 
\label{fig 1}
\vspace{.75cm}

\caption { Plots of $\varepsilon/ T^4 $ as a function of $T/T_c$ from
our model and lattice results (symbols) for gluon plasma.} 
\label{fig 2}
\vspace{.75cm}
\end{figure}
\begin{table}
\caption { Various thermodynamic quantities of ($e$, $e^+$, $\gamma$) system 
from our model with plasma and without plasma.} 
\label{Table 1}
\vspace{.75cm}
\end{table}
\vspace{5cm}
\begin{center}
{\bf Table 1} \\[1cm]
\begin{tabular}{||c|c|c|c|c|c|c||}
\hline
Medium&$f_e^2$&$n_e\,[(T/\hbar c)^3]$ &$n_{\gamma}\,[(T/\hbar c)^3]$ 
&$\varepsilon_e\,[T^4/(\hbar c)^3]$ 
&$\varepsilon_{\gamma}\,[T^4/(\hbar c)^3]$ \\
\hline
Plasma&1.79925 &0.18230 &0.24164 &0.57527 &0.65705 \\
\hline
Vacuum&1.80309 &0.18254 &0.24339 &0.57526 &0.65744 \\
\hline
\end{tabular}
\end{center}


\newpage
\begin{thebibliography}{99} 
\bibitem{me.1} M. V. Medvedev, Phys. Rev. {\bf E59}, R4766 (1999). 
\bibitem{qgp.1} Proceedings of 5$^{th}$ International Conference on 
Physics and AstroPhysics of Quark Gluon Plasma, Kolkata, India (2005), 
www.veccal.ernet.in/~icpaqgp. 
\bibitem{pe.1} A. Peshier, B. Kampfer, O. P. Pavlenko and 
G. Soff, Phys. Lett. {\bf B337}, 235 (1994). 
\bibitem{pe.2} A. Peshier, B. Kampfer, O. P. Pavlenko and 
G. Soff, Phys. Rev. {\bf D54}, 2399 (1996); 
P.Levai and U. Heinz, Phys. Rev. {\bf C57}, 1879 (1998); R. A. Schneider 
and W. Weise, Phys. Rev. {\bf C64}, 055201 (2001);  
D. H. Rischke, nucl-th/0305030, 2003.
\bibitem{go.1} M. I. Gorenstein and S. N. Yang, Phys. Rev. {\bf D52}, 
5206 (1995).  
\bibitem{ba.1} V. M. Bannur, hep-ph/0508069. 
\bibitem{pa.1} R. K. Patria, {\it Statistical Mechanics}, 
ButterworthHeinemann, Oxford (1997). 
\bibitem{ka.1} G. Boyd, J. Engels, F. Karsch, E. Laermann, C. Legeland, 
M. Lutgemeier and B. Petersson, Phys. Rev. Lett. {\bf 75}, 4169 (1995);  
Nucl. Phys. {\bf B469}, 419 (1996); F. Karsch, Nucl. Phys. A, {\bf 698}, 
199 (2002); E. Laermann and O. Philipsen, Ann. Rev. Nucl. Part. Sci. {\bf 53}, 
163 (2003). 
\end{thebibliography}
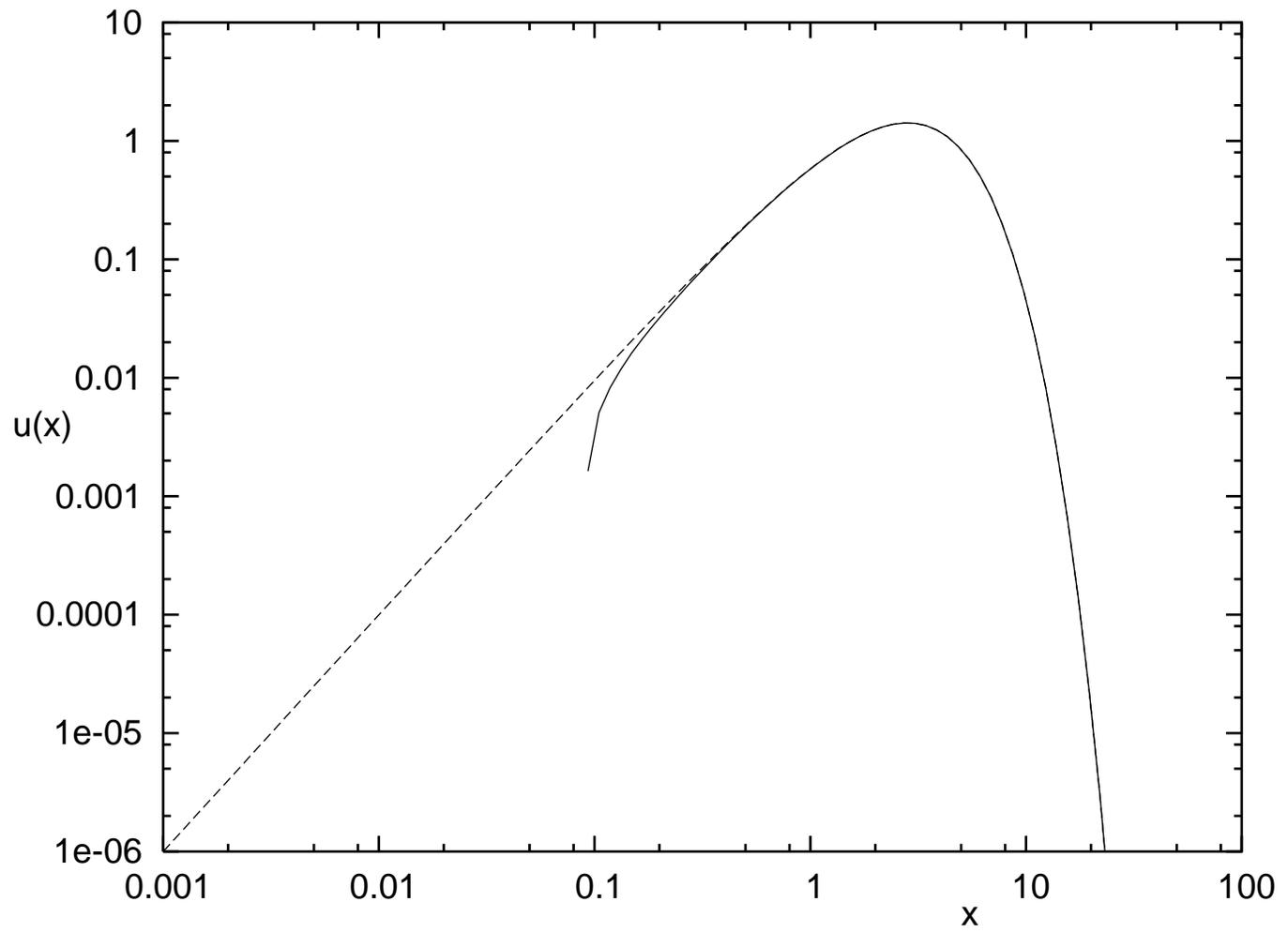
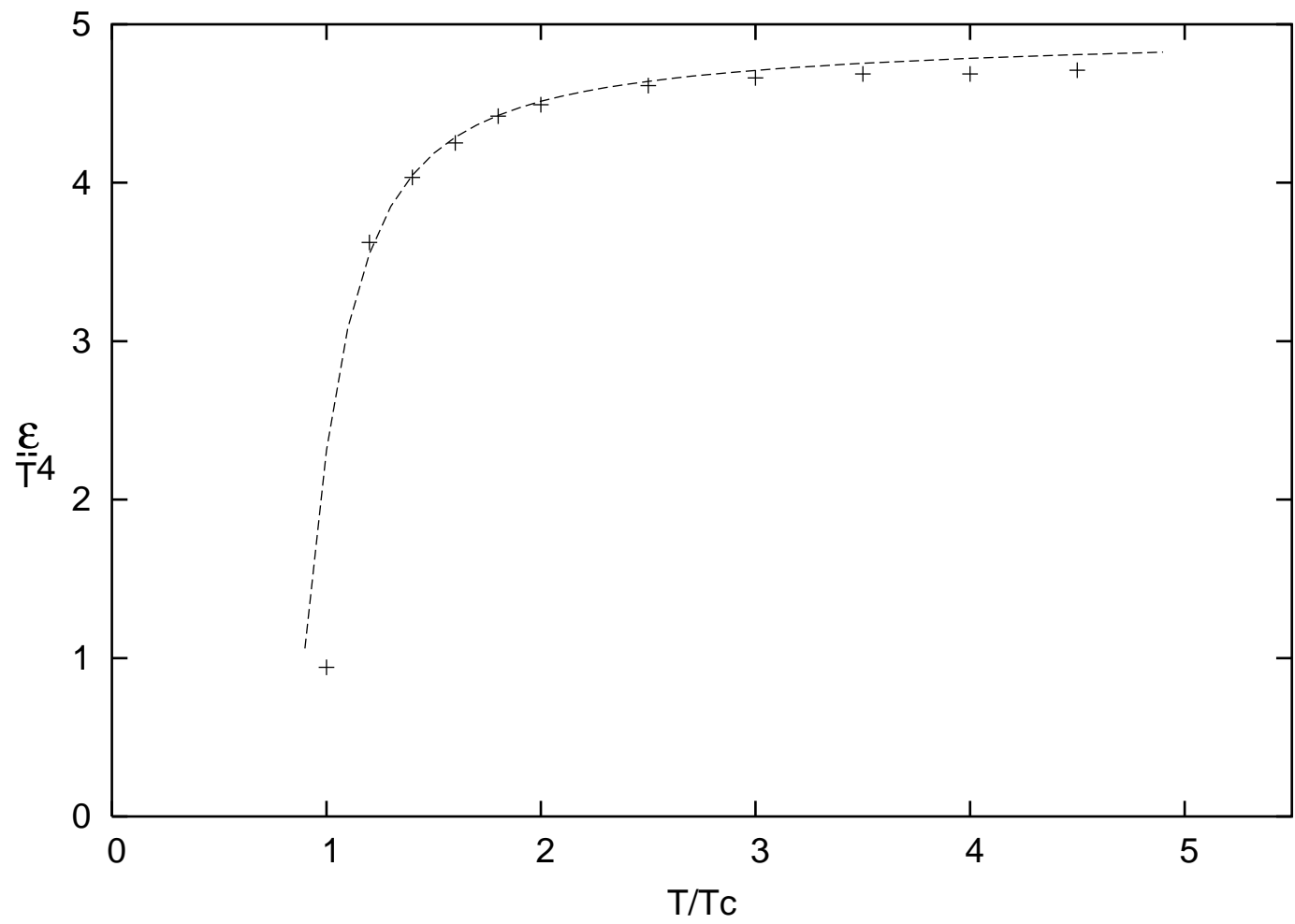
\end{document}